\documentclass[prl,showpacs,twocolumn]{revtex4}
\usepackage{graphicx}

\begin{document}

\title{EVIDENCE FOR NEARBY SUPERNOVA EXPLOSIONS}

\author{Narciso Ben\'\i tez}
\affiliation{
Department of Physics and Astronomy, Johns Hopkins University, 3400
N. Charles St., Baltimore, MD 21218, USA; e-mail: txitxo@pha.jhu.edu
}

\author{Jes\'{u}s Ma\'{\i}z-Apell\'{a}niz}
\affiliation{
Space Telescope Science Institute, 3700 San Martin Drive, Baltimore,
MD 21218, USA; e-mail: jmaiz@stsci.edu
}
\author{Matilde Canelles}
\affiliation{
Summit Hills, 1705 E.West Hwy, Silver Spring, MD, 20910, USA;
e-mail: mcanelles@niaid.nih.gov
}

\date{\today}

\begin{abstract}
Supernova explosions are one of the most energetic---and potentially
lethal---phenomena in the Universe. Scientists have speculated for
decades about the possible consequences for life on Earth of a nearby
supernova, but plausible candidates for such an event were lacking.
Here we show that the Scorpius-Centaurus OB association, a group of
young stars currently located at $\sim$ 130 parsecs from the Sun,
has generated 20 SN explosions during the last 11 Myr, some of them
probably as close as 40 pc to our planet. We find that the deposition
on Earth of \( ^{60} \)Fe atoms produced by these explosions can
explain the recent measurements of an excess of this isotope in deep
ocean crust samples. We propose that $\sim$ 2 Myr ago, one of the SNe
exploded close enough to Earth to seriously damage the ozone layer,
provoking or contributing to the Pliocene-Pleistocene boundary marine
extinction.
\end{abstract}


\maketitle


It has been proposed that the Local Bubble, a \( 150 \) pc
hot (\( T\approx 10^{6} \)K), low-density gas cavity which surrounds the
solar system, was formed by several SN explosions during the last
\( \approx 10 \) Myr \cite{SC01}. The paucity of
SNe in the Galaxy makes very unlikely that several isolated SN explosions
would happen in short succession within such a small region, but about
\( 20\% \) of all SNe originate in OB star associations, and
are therefore strongly clustered in time and space. By tracing back in time
the positions of all nearby OB associations, it is possible to
show that the Sco-Cen association was the only one able to produce SNe in the
right numbers and places to generate the Local Bubble \cite{MA01}.

\begin{figure}[tb]
\includegraphics[width=8.cm]{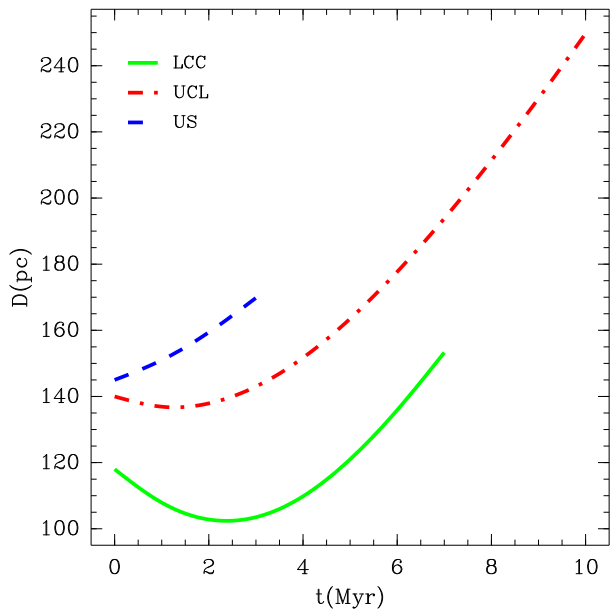}
\caption{
Evolution of the total distance between the Sun and Sco-Cen subgroups
during the last 11 Myr. For each subgroup, only the epoch during which
SNe were being formed is shown (see also Fig. 1 of Ref. \cite{MA01} which
shows
the projection onto the Galactic Plane of the positions of the Earth and the
Sco-Cen association)
}
\end{figure}

This association can be divided into three subgroups \cite{dZ99}:
Lower Centaurus Crux (LCC), Upper Centaurus Lupus (UCL) and Upper
Scorpius (US). 
Using detailed age \cite{dG89} and membership information \cite{dZ99}, it is
possible to
compare the current numbers of early/late OB stars in each subgroup with the
predictions
of synthetic star formation models. The
agreement with these
models is good and based on them it can be
estimated that each subgroup
started
producing SNe at 3-5 Myr after their formation, i.e. \( 7-8 \), \( 10-11 \)
and \( 2-3 \)
Myr ago respectively, with an expected constant rate of  \( \sim 1 \) SN
Myr\( ^{-1} \) (see Ref. \cite{MA01} for details).

The current distances to the Sco-Cen subgroups can be reliably calculated
using trigonometric
parallaxes \cite{MA01a}, and then traced back in time taking into account the
motion of both the
Sco-Cen association and the Sun with respect to the local standard of rest
which rotates with
the Galaxy, as described in Ref. \cite{MA01}. Fig 1. shows how the distance
from Earth of each
subgroup has evolved during the epoch in which they were actively
producing SNe.  At its closest, about \( 2-3 \) Myr ago, the center of
LCC was at \( \sim 100 \) pc from the Solar System; the spatial extent
of these groups can be approximated by a Gaussian with $\sigma=25-30$
pc, what means that SNe from LCC could have exploded as close as \(
\sim 40 \) pc from Earth (\( 2\sigma \) lower limit).


The explosion of a nearby SN can be detected by isotope anomalies 
in the geological record caused by the deposition of SN 
debris \cite{EFS96}. Recently, Knie et al. \cite{K99} 
measured a significant excess of \( ^{60} \)Fe atoms
in two layers of deep ocean crust corresponding
to the intervals \( 0-2.8 \), \( 3.7-5.9 \) Myr (see Table 1). They
concluded that these \( ^{60} \)Fe atoms could only be produced by a SN explosion, which 
they proposed took place about \( \sim 5 \) Myr ago at \( D\sim 30 \) pc,
causing the excess of \( ^{60} \)Fe in the second layer. The youngest layer results 
were tentatively explained as due to a background of
radioactive iron in the solar neighborhood. A reanalysis of these data came to similar conclusions\cite{FE99}, 
but attributed the presence of $^{60}$Fe in the younger layer to biomixing. 
Here we propose that the origin of the $^{60}$Fe atoms are the Sco-Cen SNe. 
As we show below, both the amplitude and timing of 
their expected deposition rate are in the right range to explain the observed excess. 

Gas ejecta from a SN explosion cannot easily 
reach the Earth unless the pressure from 
the SN blast front is larger than the ram pressure of the solar wind at the Earth 
orbit. For an isolated SN, whose front reaches the Solar System driven by momentum 
conservation\cite{mc77}, this roughly defines a cut-off distance of $15-100$ pc 
(depending on the geometrical configuration of the SN, the Sun and the Earth)\cite{EFS96}. 
This is also true in the case of multiple SN, though in that case 
the second and subsequent SNe explode in a very low density medium, 
making their expansion follow the Sedov regime\cite{bl97,sed,SC01} 
up to distances comparable with Local Bubble size $\sim 200$ pc. 
Although this would make it
difficult for {\it gaseous} debris from all but the closest Sco-Cen SNe 
to penetrate the heliosphere up to the Earth's radius, observations show that most of the iron 
in the Local Bubble is condensed forming dust \cite{bl86,SC01}. The interaction of interstellar dust 
with the solar wind and magnetic field is a complex problem, which 
strongly depends on the size of the dust particles\cite{ra}, but 
interstellar dust containing iron has been found at the Earth's orbit 
\cite{gr94}, and it seems reasonable to assume that most, or at least a large fraction 
of the iron reaching the heliosphere traveling in a SN blast front would have 
reached the Earth's orbit.\cite{ra}.

Table 1 presents the typical distance and expected number of Sco-Cen 
SNe in each of the corresponding time intervals. The 
expected surface density of \( ^{60} \)Fe (corrected for
in situ decay) deposited in a layer \( N(\Delta l) \)
can be estimated \cite{FE99} as:

\[
N(\Delta l)=4.1\times 10^{7}N_{\rm {SN}}f
\left( \frac{M_{^{60}{\rm{Fe}}}}{10^{-5}\mbox {M}_{\odot }}\right)
\left( \frac{100\, {\rm{pc}}}{D}\right) ^{2}\mbox {cm}^{-2}\; ,\]

\noindent where \( N_{\rm {SN}} \) is the number of SNe, which are assumed to
happen at a typical distance $\sim D(\Delta l)$ during the time interval
covered
by each of the sediment layers, \( f \)
is the uptake factor that \cite{K99} estimate as \( 1/100 \), and
\( M_{^{60_{\rm Fe}}} \)is the expected \( ^{60} \)Fe yield by a SN.
To compare with the results of Ref. \cite{K99}, we divide \( N(\Delta l) \)
by the \( \Delta t \) covered by each layer, which yields the flux
\( \phi _{\rm {SN}} \) (cm\( ^{-2} \) Myr \( ^{-1} \))presented in Table 1.

\begin{table}
\caption{\sc Predictions and measurements of \( ^{60} \)Fe excess in
deep oceanic crust samples (corrected for in situ decay)}
\begin{center}
\begin{tabular}{|c|ccc|}
\hline
  & Layer 1&  Layer 2& Layer 3\\
\hline
age(Myr)&  0-2.8&  3.7-5.9&  5.9-13                                  \\
\hline
\( N_{\rm SN} \)&  8&  4&  6                                         \\
\hline
D\( _{\rm SN} \) (pc)& 130&  140&  205                               \\
\hline
\( \phi_{\rm SN} \) (10\( ^{6} \)cm\( ^{-2} \) Myr\( ^{-1} \))&
\( 0.7_{-0.06}^{+6.30} \)&
\( 0.4_{-0.04}^{+3.6} \)&
\( 0.08_{-0.01}^{+0.8} \)                                            \\
\hline
\( \phi_{\rm b} \)(10\( ^{6} \) cm\( ^{-2} \) Myr\( ^{-1} \))&
\( 0.1 \)1&
\( 1.5 \)&
\( 5 \)                                                              \\
\hline
\( \phi _{\rm SN}+\phi _{\rm b} \) (10\( ^{6} \)cm\( ^{-2} \) Myr\( ^{-1}
\))&
\( 0.81_{-0.06}^{+6.30} \)&
\( 1.9_{-0.04}^{+3.6} \)&
\( 5.08_{-0.01}^{+0.8} \)                                            \\
\hline
\( \phi _{\rm obs} \) (10\( ^{6} \) cm\( ^{-2} \) Myr\( ^{-1} \))&
\( 1.0_{-0.3}^{+0.5} \)&
\( 8_{-5}^{+11} \)&
\( 10_{-8.5}^{+22} \)                                                \\
\hline
\end{tabular}
\end{center}
\end{table}

This estimation has several sources of uncertainty. The
\( ^{60} \)Fe yield can vary from \( 10^{-4} \) M\( _{\odot } \)
to \( 10^{-6} \) M\( _{\odot } \) depending on the SN mass (type
II SN, \cite{WW95,timmes,migue,plues1,plues2,migue2}). Also, the uptake
factor \( f \),
which represents the fraction of \( ^{60} \)Fe
atoms present in the ocean which is deposited in the crust, has large
and difficult to estimate uncertainties \cite{FE99}.
To compound these two error
sources, the positions and time of the explosions can have a considerable
scatter, and even the total number of SNe has a Poisson uncertainty of $
 24\% $. Therefore, we consider the values of \( \phi _{\rm SN} \)
in Table 1 to be, at best, order-of-magnitude estimates, and that
is reflected in the error bars assigned to our predictions in Fig
2. Table 1 also presents the inferred average fluxes
\( \phi _{\rm obs} \) measured by Ref. \cite{K99}, and the expected
background level \( \phi _{\rm b} \)
based on a measurement of a \( 13 \) Myr old core. Note that in Fig
2, we have added the \( ^{60} \)Fe background rate \cite{K99}
\( \phi _{\rm b} \) to the contribution from the Sco-Cen SNe
\( \phi _{\rm SN} \), and that the error bars correspond only to the
uncertainty in \( \phi _{\rm SN} \), since the above reference does not
provide an error estimate for the background.

The agreement is excellent for the first and youngest layer, which has
the highest signal-to-noise. The flux measured in the second layer
is a factor 4 higher than our prediction, but there is an ample overlap
between the error bars of both quantities, and thus the results can
be considered consistent. Regarding the third, oldest layer,
it is not clear whether there is any signal above the background,
so our prediction of a very small flux is also compatible with the
observation . It should also be noted that no signal of SN origin was detected in the same layers for
another isotope $^{53}$Mn \cite{K99,FE99}, which is again consistent with our 
scenario since the predicted $^{53}$Mn deposition rates are much lower 
than the background of cosmogenic origin. 
We therefore conclude that the Sco-Cen SNe are enough to explain the excess 
of $^{60}$Fe in the deep ocean crust. It will be very interesting to obtain
crust data with better ``temporal'' resolution to identify individual SN explosions;
including in the search other SN-created radioisotopes with low backgrounds and proper
decay rates would allow to pin down the mass range of the progenitors and the
distance at which the SNe exploded.

\begin{figure}[tb]
\includegraphics[width=8.cm]{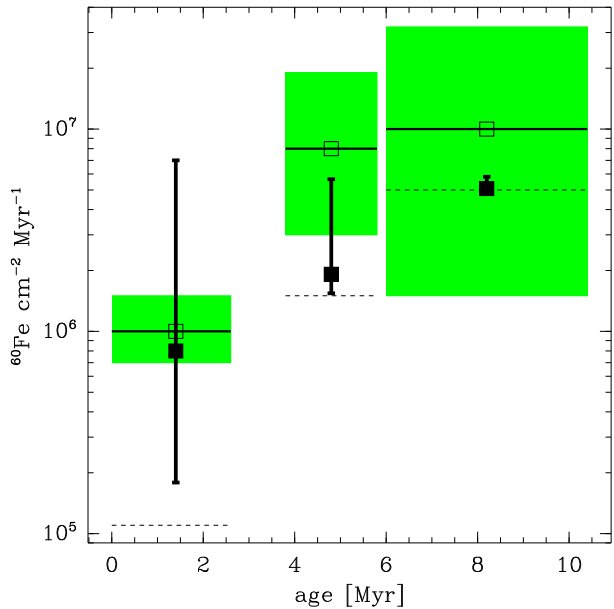}
\caption{Comparison between the \( ^{60}\rm {Fe} \) deposition rate
predicted from the Sco-Cen SNe and the measurements of deep ocean
crust samples. The horizontal continuous lines/empty squares and the shadowed boxes
represent respectively the data and errors of Ref. \cite{K99}, while our
predictions and associated error estimates are represented by the 
filled squares and error bars. The dashed lines correspond to the background
estimates of Ref. \cite{K99}}
\end{figure}


Several authors \cite{KS57,S68,RU71,RU74} have proposed that the explosion
of a nearby SN could have caused one or more of the massive extinction
events in the fossil record. At distances larger than a few pc,
the only component of the SN emission capable of performing serious damage
to the biosphere is the charged cosmic ray radiation \cite{ES95} (for
instance, hard UV radiation from the SN shock break-out would produce flux levels 
above the atmosphere of $F_{UV}\sim 10^7 (50 \rm{pc}/D)^2$ ergs cm$^-2$ during 1
day \cite{sca}, a level smaller than the amount of similar radiation received from the
Sun \cite{C00}). A strong increase in the flux of cosmic rays reaching the upper levels of
the atmosphere speeds up the production of NO, which catalytically destroys large amounts
of ozone molecules \cite{RU74}. Assuming that the energy released by the SN in
cosmic rays is \( 10^{50} \) ergs, the time integrated flux or fluence
reaching Earth will be \cite{ES95}:

\[
\phi _{CR}\Delta t\approx 2.2\times 10^{9}\left( \frac{10\, \mbox
{pc}}{D}\right) ^{2}
\mbox {ergs}\, \mbox {cm}^{-2}
\]

It has usually been considered in the literature \cite{L68,RU74,ES95}
that the cosmic rays would travel by diffusion in a 
random Galactic magnetic field with a scale-length of 1 pc and that,
therefore, the above flux would be spread over a period of
\( \Delta t\approx 3(D/1\mbox{ pc})^{2} \) yr. However, the assumption of
randomness is not valid in this case: the magnetic field in the outer
'shell' of
the Local Bubble is probably coherent on large scales and
very weak or inexistent in its interior \cite{H98}, being pushed out
by the effects of the SN explosions.
Therefore, the amount of cosmic rays reaching the Earth in the event of a
nearby SN explosion
would strongly depend on the relative position of the
SN with respect to Earth and the Local Bubble shell. If both the SN
and the Sun are within the cavity, the cosmic rays will probably travel
unhindered and hit the Earth spread over a $ \Delta t \sim \tau$, where
$\tau$ corresponds to their emission period, estimated to be in the range $
 \tau \sim 10-10^{5} $yr
\cite{ES95,key-21}. If the SN were in the outside, the Bubble shell
would serve as a \char`\"{}conductor\char`\"{} for the cosmic
rays if the Sun were also outside the Bubble, or as an {}``insulating{}''
shield, if the Sun were inside.

This means that the value of the time interval \( \Delta t \) is highly
uncertain, and the best we can do is estimate the \emph{maximal} CR flux
produced
by one of the Sco-Cen SNe by taking the fluence above and a \( \Delta t \)
similar to the lower limit of the typical cosmic ray acceleration
time \( \tau \sim 10 \) yr. Using the minimal distance for the Sco-Cen
SNe \( D\geq 40 \)pc we find that \( \phi _{CR}\leq 1.4\times 10^{7} \)ergs
cm\( ^{-2} \)yr\( ^{-1} \), within a factor 2 of the flux estimated
by \cite{ES95} for a \( 10 \) pc event using the random magnetic
field assumption. It has been estimated \cite{key-23}
that such an event would lead to a ozone depletion of \( 60\% \) at
high latitudes and about \( 20\% \) at the equator.

Therefore, the subsequent increase in the UV-B flux from the Sun 
reaching the Earth surface due to one or more of the Sco-Cen SNe could 
have caused at most a minor extinction \cite{S86,C99}, but would not be enough 
to provoke a major mass extinction like e.g. the Cretaceous-Tertiary event
(see also \cite{sca}, where the mutagenic effects of excess UV radiation
are discussed in detail). This extinction would particularly affect marine
ecosystems, as first proposed by Ref. \cite{ES95}; 
an increase on the UV-B flux over the ambient level can provoke a 
significant reduction in phytoplankton abundance and biomass, propagating 
to at least one species of zooplankton as a secondary effect \cite{K97}. 
Despite the uneven distribution of the ozone depletion, tropical species 
would be more affected due to the higher solar angle at which they receive their
UV dose \cite{C99}. Therefore, the biological signature of a SN explosion
provoking a ``clean'' UV-B catastrophe would be a decline of ocean surface
phytoplankton productivity not associated with other causes as volcanic
activity, climate changes or impact events. A decline in plankton productivity 
would be very difficult to detect in the fossil record, but it could be 
inferred from secondary effects, in particular an extinction of mollusks \cite{R96}.

Such an extinction affecting marine tropical, subtropical and temperate 
American bivalves characterized the Pliocene-Pleistocene boundary, \( \sim 2 \) 
Myr ago \cite{S81,S86,V86,P95} (see also Ref. \cite{berk}). 
Two hypotheses have been advanced to explain this phenomenon: the emergence 
of the Panama isthmus \cite{A93,A01} and cooling due to the onset
of Northern Hemisphere glaciations \cite{ST86,J93, J94}. However, this 
extinction episode was too rapid to be due to the Panama isthmus closure 
\cite{J93,J94} and a detailed analysis of the extinction patterns seems to 
rule out cooling as the cause \cite{A93,R96}. A simultaneous, although slower, 
episode of extinction affected corals \cite{J94}, which are known to be highly 
susceptible to UV-B radiation \cite{C99}. This leaves a SN-provoked UV-B catastrophe as a possible candidate for the Pleistocene--Pliocene extinction; it should be 
noted that this epoch roughly coincides with the time of closest approach of LCC 
(see Fig. 1), during which the probability of nearby SN explosions would 
have been highest. In addition, Ref. \cite{bar} proposed that a SN blast wave 
ionized the local interstellar medium between 2 and 3.6 Myr ago.

To test this hypothesis, the time and distance at which individual SN explosions took place 
should be determined more precisely, using geological
information as suggested above.  A coincidence in time between the SN 
expected to have strongest effects on the biosphere and the
Pleistocene-Pliocene extinction, would strongly support the existence of a link 
between both events. 

\smallskip

We would like to thank Mario Livio, Santiago Arribas, Marc Davis, Carl
Heiles, Holland Ford, Martin Barstow, and Dale Russell for helpful comments. 
N.B acknowledges support 
from the NASA ACS grant and the Center for Astrophysical Sciences at JHU.
J.M.-A. acknowledges support from grant 82280 from the STScI DDRF.

\end{document}